\documentclass[cameraready]{Interspeech}
\usepackage{multirow}

\title{PhASE-Flow: Phonetic-Conditioned Acoustic Flow Matching in SSL Representation Domain for Speech Enhancement}

\author[affiliation={1,2}, orcid=0009-0000-2417-9904]{Jun}{Gao}
\author[affiliation={1,2}, orcid=0009-0006-4373-8306]{Xiaobin}{Rong}
\author[affiliation={3}]{Yu}{Sun}
\author[affiliation={1,2}, orcid=0009-0003-8267-6994]{Dahan}{Wang}
\author[affiliation={1,2}, correspondingauthor]{Jing}{Lu}

\address{
    $^1$ Key Laboratory of Modern Acoustics, Nanjing University, Nanjing, China \\
    $^2$ NJU-Horizon Intelligent Audio Lab, Horizon Robotics, Beijing, China \\
    $^3$ R\&D Centre, Samsung Electronics (China), Nanjing 210019, China
}

\email{\{jun.gao, xiaobin.rong, dahan.wang\}@smail.nju.edu.cn, yuyu.sun@samsung.com, lujing@nju.edu.cn}


\usepackage{comment}

\begin{document}

\maketitle

\begin{abstract}
    Flow matching (FM) enables high-fidelity generation, while self-supervised learning (SSL) speech models provide hierarchical representations spanning acoustic and phonetic levels. However, existing FM-based speech enhancement (SE) methods operate primarily in the spectral domain, treating SSL features only as external conditions rather than modeling directly in the SSL latent space. To fully exploit the structural richness of SSL representations, we propose PhASE-Flow, an FM-based SE framework that operates entirely in the SSL space. It models the conditional distribution of clean acoustic representations given phonetic ones, reconstructing the waveform via a neural vocoder. Experiments show that PhASE-Flow outperforms state-of-the-art baselines in perceptual quality and intelligibility. Notably, it achieves competitive performance with only four sampling steps, enabling highly efficient inference. Audio demos are available at \urlstyle{same}\url{https://anonymous.4open.science/w/phase-flow_demo-E6E1/}.
\end{abstract}

\section{Introduction}

Speech enhancement (SE) aims at recovering clean speech from noisy observations to improve perceptual quality and speech intelligibility. While conventional discriminative methods are effective at noise attenuation, they often struggle to preserve speech naturalness under challenging acoustic conditions~\cite{wang25s_interspeech}.

Recently, generative methods have emerged as a powerful paradigm for SE, driven by their exceptional capability to synthesize speech with high perceptual quality~\cite{SenSE}. By modeling the underlying distribution of clean speech, these models reconstruct the target signal via conditional generative formulations, including generative adversarial networks (GANs)~\cite{sun25g_interspeech}, diffusion models~\cite{SGMSE,StoRM}, flow matching (FM)~\cite{wang25s_interspeech}, and language models (LMs)~\cite{SELM,Genhancer}.
Among these, LM-based SE approaches that utilize discrete representations inevitably discard fine-grained acoustic details, often resulting in compromised fidelity~\cite{PASE}. Additionally, GANs are typically hindered by training instability and mode collapse. In contrast, diffusion- and FM-based methods have demonstrated superior quality, despite requiring iterative sampling that may limit inference efficiency~\cite{rethinkingflow}.

Existing FM-based SE methods predominantly model the clean speech distribution in the spectral domain, as it preserves rich acoustic information for waveform reconstruction. Specifically, methods like FlowSE~\cite{wang25s_interspeech} generate clean Mel spectrograms from noisy inputs, followed by neural vocoder synthesis. In contrast, approaches like~\cite{SGMSE,StoRM,pretrainSR} operate in the short-time Fourier transform (STFT) domain, enabling direct waveform reconstruction via inverse STFT (iSTFT) without an additional vocoder. To further improve linguistic integrity, recent studies~\cite{wang25s_interspeech, SenSE} have conditioned Mel-domain generation on semantic\footnote{Semantic is sometimes referred to as \emph{phonetic}, as the underlying representations are closer to phoneme-level units.} prompts, typically derived from high-level self-supervised learning (SSL) speech representations, leading to improved content accuracy.

Despite the success of semantic conditioning, current methods mostly rely on the straightforward integration of semantic cues with spectral features, leaving the optimal representation domain unexplored. We contend that both Mel and STFT domains suffer from inherent limitations. While Mel spectrograms offer compact, perceptually aligned representations, they lack explicit phase information, thereby limiting the upper bound on reconstruction quality. Conversely, STFT preserves phase but exhibits heavy-tailed distributions~\cite{gerkmann2010empirical}, which complicate statistical modeling and may reduce reconstruction stability.
More fundamentally, spectral features are low-level representations where pitch, timbre, and linguistic content are tightly entangled, which may make generative modeling inherently difficult.

In contrast to spectral features, SSL representations exhibit a disentangled hierarchical structure, where lower layers capture acoustic details and higher layers encode phonetic properties~\cite{LayerWiseS3M}, and can thus be categorized into \emph{acoustic} and \emph{phonetic} types~\cite{PASE}.
Recent work~\cite{Investigating_SSL_for_SE, Boosting_SSL_SE, sun25g_interspeech} has demonstrated that acoustic SSL representations yield superior perceptual quality over traditional spectral features in SE, indicating that they contain richer and more structured acoustic details. Given that acoustic SSL representations retain fine-grained acoustic details while phonetic ones provide semantic cues, jointly modeling them could offer a unified representation space that aligns semantic content with acoustic structure. Such alignment reduces representational mismatch and potentially provides a more structured and semantically coherent foundation for FM-based generative SE.

\begin{figure*}[t]
    \centering
    \includegraphics[width=0.9\linewidth]{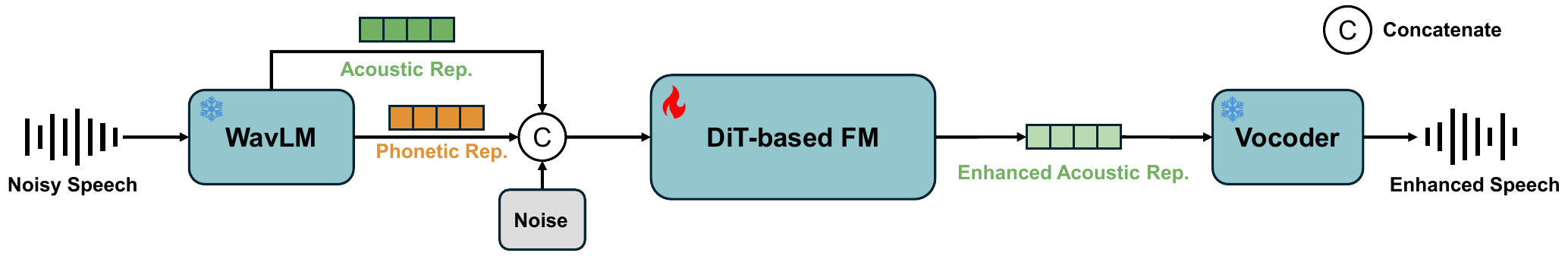}
    \caption{Overview of the proposed PhASE-Flow framework.}
    \label{fig:PhASE-Flow}
\end{figure*}

Building on this idea, we propose \textbf{PhASE-Flow}, a \textbf{Ph}onetic-conditioned \textbf{A}coustic \textbf{S}peech \textbf{E}nhancement framework using \textbf{Flow} matching. This framework operates entirely within the SSL representation domain derived from WavLM~\cite{WavLM}. Similar to~\cite{PASE}, PhASE-Flow leverages two types of WavLM representations: (1) acoustic representations extracted from the first Transformer layer, and (2) phonetic representations from the final Transformer layer. Our key design is to employ a DiT-based~\cite{DiT} FM module to model the manifold of acoustic representations conditioned on phonetic ones.
During inference, the generated representations are converted into enhanced waveforms using a pre-trained neural vocoder. Experimental results demonstrate that PhASE-Flow achieves substantial improvements in speech quality, intelligibility, and speaker similarity. Notably, our framework delivers competitive performance with only four sampling steps, significantly reducing inference latency compared to diffusion-based methods \cite{SGMSE,StoRM}.

\section{Method}

\subsection{Framework Overview}

As illustrated in Figure~\ref{fig:PhASE-Flow}, PhASE-Flow comprises three integral modules: (1) a frozen WavLM encoder to extract acoustic and phonetic representations from noisy inputs; (2) a trainable DiT-based FM module, whose backbone is adapted from~\cite{f5tts}, to model the distribution of clean acoustic representations; and (3) a pre-trained neural vocoder for waveform reconstruction.

\subsection{WavLM Encoder}

WavLM has established itself as a cornerstone of SSL speech models, encoding acoustic and phonetic information within its lower and higher Transformer layers~\cite{LayerWiseS3M}, respectively. However, existing SSL-based SE methods face two major limitations. First, they primarily rely on deterministic noisy-to-clean mappings~\cite{sun25g_interspeech}, which inevitably cause over-smoothing when modeling complex, multi-modal speech distributions. Second, they predominantly exploit the semantic information from higher layers, largely neglecting the well-structured acoustic details preserved in lower layers. Consequently, the optimal strategy for leveraging this latent space in a generative paradigm remains under-explored.

Through preliminary evaluations of various layer configurations, we identify an effective synergy that can maximize the potential of this latent space.
Specifically, we extract two disentangled representations from WavLM: (1) the \textbf{acoustic} representation from the first Transformer layer, whose manifold is modeled by the flow, and (2) the \textbf{phonetic} representation from the final Transformer layer that acts as the condition for generative modeling.
While this layer disentanglement aligns with findings in~\cite{PASE}, our work fundamentally differs by exploring its potential within a continuous-space FM framework.

Let $y \in \mathbb{R}^T$ be the input raw waveform. The acoustic and phonetic representations from WavLM are denoted as $z_a, z_p \in \mathbb{R}^{N \times D}$, where $N$ is the number of frames and $D$ is the feature dimension.

\subsection{Phonetic-Conditioned Acoustic Flow Matching}

Modeling dense, well-structured manifolds (e.g., SSL speech representations) via score-based diffusion models is inherently inefficient~\cite{li2025back}, and estimating intricate score functions forces the use of fine-grained steps during sampling~\cite{lipman2023flow}.
Conversely, advanced frameworks like Schr\"{o}dinger bridge and flow matching can leverage parameterization tricks to directly predict clean targets, enabling highly effective manifold modeling. To achieve a favorable balance between generation quality and sampling efficiency, PhASE-Flow formulates its generative process as an FM-based ordinary differential equation (ODE)~\cite{lipman2023flow}:
\begin{equation}
    \frac{\mathrm{d} z_t}{\mathrm{d} t} = u_t(z_t \mid z_{a,s}),
\end{equation}
where $u_t$ denotes the time-dependent conditional vector field for $t \in [0,1]$. The conditioning term $z_{a,s}$ represents the acoustic representation, with the subscript $s$ indicating that it is derived from the clean signal.

Restricting $z_t$ to follow a Gaussian probability path $p_t (z_t \mid z_{a,s}) = \mathcal{N} (\mu_t, \sigma_t^2 I)$, we use the widely adopted optimal transport (OT) conditional vector field~\cite{lipman2023flow}, defined as $\mu_t = t z_{a,s}$ and $\sigma_t = 1 - t$.
The intermediate state $z_t$ can be constructed as $z_t = \mu_t + \sigma_t z_0$, where $z_0 \sim \mathcal{N}(0,I)$ represents Gaussian noise sampled from the prior distribution, and $t \sim \mathcal{U}(0,1)$ denotes the sampled flow step.

Among the feasible prediction targets, including data ($x$-pred), vector field ($v$-pred), and noise ($\epsilon$-pred) prediction, we observe that $x$-pred yields the most stable loss convergence and superior performance, consistent with~\cite{li2025back}. Consequently, we adopt $x$-pred as the training objective:
\begin{equation}
\mathcal{L}(\theta) = \mathbb{E}_{t, z_{a,s}, z_t} \lVert x_{\theta}(z_t, t, z_{p,y}, z_{a,y}) - z_{a,s} \rVert^2,
\end{equation}
where $z_{p,y}$ and $z_{a,y}$ denote the phonetic and acoustic representations extracted from the noisy utterance paired with the clean target. The subscript $y$ refers to the noisy signal.

The DiT-based backbone receives four inputs: (1) a noisy acoustic representation~$z_{a,y}$, (2) a noisy phonetic representation~$z_{p,y}$, (3) an intermediate state~$z_t$, and (4) a sampled flow step~$t$. To ensure effective utilization of the phonetic representation, we randomly drop the acoustic representation~$z_{a,y}$ with probability~$p_a$ during training.
Ablation studies demonstrate that noisy phonetic representations effectively facilitate the modeling of acoustic representations.

During inference, we derive the velocity vector field $v_{\theta}$ from the predicted data $x_{\theta}$ using the relationship $v_{\theta} = (x_{\theta} - z_t) / (1 - t)$. We then generate the samples by solving the ODE via Euler discretization:
\begin{equation}
z_{t + \Delta t} = z_t + v_{\theta} (z_t, t, z_{p,y}, z_{a,y}) \Delta t,
\end{equation}
where $\Delta t$ is the step size.

\subsection{Vocoder}

The vocoder reconstructs the waveform from the enhanced acoustic representations. It is trained independently on clean speech and integrated without joint fine-tuning. The vocoder uses the improved Vocos backbone~\cite{siuzdak2023vocos} introduced in~\cite{ji2025wavtokenizer}, which incorporates an attention module to enhance contextual modeling. Following~\cite{PASE}, training combines a multi-scale Mel-spectrogram reconstruction loss with adversarial and feature-matching losses from a multi-period discriminator (MPD) and a multi-band multi-scale STFT discriminator (MBMSD)~\cite{HiFi-IRVQGAN}.

\begin{table}[t]
\centering
\caption{Vocoder performance comparison on the clean speech of DNS 2020 no-reverb test set.}
\setlength{\tabcolsep}{0.8mm}
\resizebox{\linewidth}{!}{
    \begin{tabular}{ccccccc}
    \toprule
    Model & DNSMOS $\uparrow$ & UTMOS $\uparrow$ & SBS $\uparrow$ & LPS $\uparrow$ & SpkSim $\uparrow$ & dWER (\%)$\downarrow$ \\ \midrule
    Clean & 3.28 & 4.14 & 1.00 & 1.00 & 1.00 & 1.00 \\ \midrule
    Vocoder-M & 3.38 & 3.85 &\textbf{0.97} & \textbf{0.98} & 0.96 & \textbf{0.98} \\
    Vocoder-A & 3.37 & \textbf{4.01} & 0.95 & 0.97 & \textbf{0.99} & 1.02 \\
    Vocoder-P & \textbf{3.39} & 3.89 & 0.95 & 0.97 & 0.65 & 2.07 \\ \bottomrule
    \end{tabular}
}
\label{tab:ablation_vocoder}
\end{table}

\begin{table}[t]
\centering
\caption{Ablation results on DNS 2020 no-reverb test set.}
\setlength{\tabcolsep}{0.8mm}
\resizebox{\linewidth}{!}{
    \begin{tabular}{ccccccc}
    \toprule
    Model & DNSMOS $\uparrow$ & UTMOS $\uparrow$ & SBS $\uparrow$ & LPS $\uparrow$ & SpkSim $\uparrow$ & dWER (\%)$\downarrow$ \\ \midrule
    Noisy & 2.48 & 2.36 & 0.80 & 0.90 & 0.96 & 3.51 \\
    Clean & 3.28 & 4.14 & 1.00 & 1.00 & 1.00 & 0.00 \\ \midrule
    Flow-S & 3.22 & 3.56 & 0.87 & 0.94 & 0.91 & 4.73 \\
    Flow-M & 3.38 & 3.76 & 0.90 & 0.94 & 0.89 & 4.65 \\
    Flow-A & 3.38 & 4.02 & 0.91 & 0.95 & 0.94 & 4.48 \\
    Flow-P & \textbf{3.41} & 4.01 & 0.89 & 0.95 & 0.52 & 4.23 \\ \midrule
    Flow-M-P & 3.33 & 3.44 & 0.91 & 0.96 & 0.89 & 3.04 \\
    PhASE-Flow & 3.40 & \textbf{4.11} & \textbf{0.93} & \textbf{0.97} & \textbf{0.94} & \textbf{2.79} \\ \bottomrule
    \end{tabular}
}
\label{tab:ablation_domain}
\end{table}

\begin{table*}[t]
\centering
\caption{Comparison results on the DNS 2020 test set.}
\setlength{\tabcolsep}{1.5mm}
\resizebox{\linewidth}{!}{
\begin{tabular}{@{}cccccccccccccc@{}}
\toprule
\multirow{2}{*}{Model} & \multicolumn{6}{c}{No-Reverb} & & \multicolumn{6}{c}{With-Reverb} \\ \cmidrule(lr){2-7} \cmidrule(l){9-14} 
 & DNSMOS $\uparrow$ & UTMOS $\uparrow$ & SBS $\uparrow$ & LPS $\uparrow$ & SpkSim $\uparrow$ & dWER (\%) $\downarrow$ &  & DNSMOS $\uparrow$ & UTMOS $\uparrow$ & SBS $\uparrow$ & LPS $\uparrow$& SpkSim $\uparrow$ & dWER (\%) $\downarrow$\\ \midrule
Noisy & 2.48 & 2.36 & 0.80 & 0.90 & 0.96 & 3.51 &  & 1.39 & 1.30 & 0.61 & 0.63 & 0.79 & 10.23 \\
Clean & 3.28 & 4.14 & 1.00 & 1.00 & 1.00 & 0.00 &  & 3.28 & 4.14 & 1.00 & 1.00 & 1.00 & 0.00 \\ \midrule
TF-GridNet & 3.34 & 3.86 & 0.91 & \textbf{0.97} & \textbf{0.96} & 2.86 &  & 2.63 & 1.42 & 0.77 & 0.88 & \textbf{0.80} & \textbf{8.86} \\
StoRM & 3.31 & 3.73 & 0.89 & 0.95 & 0.95 & 4.41 &  & 2.87 & 1.84 & 0.61 & 0.60 & 0.55 & 49.65 \\
LLaSE-G1 & \textbf{3.42} & 3.84 & 0.84 & 0.90 & 0.74 & 12.15 &  & 3.35 & 2.90 & 0.71 & 0.70 & 0.44 & 41.66 \\
AnyEnhance & \textbf{3.42} & 3.96 & 0.91 & 0.96 & 0.91 & 4.58 &  & 3.20 & 2.75 & 0.80 & 0.87 & 0.72 & 14.16 \\
FlowSE & 3.38 & 3.76 & 0.90 & 0.94 & 0.89 & 4.65 &  & 3.34 & 3.51 & 0.81 & 0.85 & 0.72 & 15.58 \\ \midrule
PhASE-Flow & 3.40 & \textbf{4.11} & \textbf{0.93} & \textbf{0.97} & 0.94 & \textbf{2.79} &  & \textbf{3.36} & \textbf{3.81} & \textbf{0.85} & \textbf{0.90} & 0.75 & 13.19 \\ \bottomrule
\end{tabular}
}
\label{tab:dns1}
\end{table*}

\section{Experiments}


\subsection{Datasets}

The clean speech corpus comprises publicly available data from the DNS5 LibriVox subset~\cite{DNS5}, VCTK~\cite{VCTK}, EARS~\cite{EARS}, and LibriSpeech~\cite{Librispeech}.
To ensure high-quality training data, we apply data filtering by retaining only samples with DNSMOS scores (OVRL, SIG, BAK, and P.808) above 3.0 and UTMOS scores above 4.0. The EARS dataset is exempt from this filtering, as non-intrusive metrics are often unreliable for atypical speech characteristics (e.g., whispering). The filtered clean speech contains approximately 1,021 hours of data.

Noise samples are drawn from DNS5, WHAM!~\cite{WHAM}, FSD50K~\cite{FSD50K}, and FMA~\cite{FMA}, and room impulse responses (RIRs) are sourced from OpenSLR26 and OpenSLR28~\cite{openSLR}. Training mixtures are generated on the fly. Each clean utterance is convolved with a randomly selected RIR with a 50\% probability, then mixed with a noise clip at a signal-to-noise ratio (SNR) uniformly sampled from $[-5, 15]$~dB.

Evaluation is performed on the official Interspeech 2020 DNS Challenge synthetic test set~\cite{DNS1}, which comprises two subsets: \emph{with-reverb} and \emph{no-reverb}. Notably, when evaluating on the with-reverb set, we use the corresponding clean no-reverb utterances as references, since our task also includes dereverberation. All audio is sampled at 16 kHz.

\subsection{Evaluation Metrics}

We evaluate our method across three dimensions: (1) \textbf{Perceptual Quality}: DNSMOS~\cite{DNSMOS} and UTMOS~\cite{UTMOS}.
(2) \textbf{Representation similarity}: SpeechBERTScore (SBS)~\cite{SBS} computed with~\cite{mHuBERT-147}, Levenshtein phoneme similarity (LPS)~\cite{LPS} computed with~\cite{phoneme-recognition}, and speaker similarity (SpkSim)\footnote{Note that this WavLM-based backbone differs from the RawNet3 used in some cited works, so the SpkSim scores are not directly comparable.} evaluated with a fine-tuned WavLM-Large-based ECAPA-TDNN\footnote{\urlstyle{same}\url{https://github.com/microsoft/UniSpeech/tree/main/downstreams/speaker_verification}}.
(3) \textbf{Linguistic integrity}: word error rate (WER) computed using Whisper-Large-v3~\cite{pmlr-v202-radford23a}. As reference transcripts are unavailable, we use the transcriptions of clean speech as pseudo-references, reporting the result as dWER.

\subsection{Baselines and Ablation Study}

We benchmark PhASE-Flow against representative state-of-the-art (SOTA) methods across diverse paradigms: the discriminative TF-GridNet~\cite{TF-GridNet}, diffusion-based StoRM~\cite{StoRM}, FM-based FlowSE~\cite{wang25s_interspeech}, LM-based LLaSE-G1~\cite{LLaSE-G1}, and masked generative model (MGM)-based AnyEnhance~\cite{AnyEnhance}.

Official checkpoints are used for TF-GridNet\footnote{\urlstyle{same}\url{https://huggingface.co/kohei0209/tfgridnet_urgent25}}, StoRM\footnote{\urlstyle{same}\url{https://github.com/sp-uhh/storm}}, and LLaSE-G1\footnote{\urlstyle{same}\url{https://huggingface.co/ASLP-lab/LLaSE-G1}}.
Since StoRM provides separate checkpoints for denoising (WSJ+CHiME3) and dereverberation (WSJ+Reverb), we sequentially apply dereverberation followed by denoising during inference. AnyEnhance is evaluated using the inference results provided by the authors.
For FlowSE, as its released checkpoint is trained on a relatively small dataset, we retrain it following the official implementation~\cite{wang25s_interspeech}. Specifically, we omit the text-conditioning modules for a fair comparison. Our reproduction achieves DNSMOS scores within 3\% of the results reported in the original paper.

To evaluate the effects of operating entirely within the SSL domain and the impact of phonetic conditioning, we design the following ablation variants:
\begin{itemize}
    \item \textbf{Flow-M}: A Mel-domain baseline relying exclusively on Mel-spectrogram inputs, implemented as the reproduced FlowSE.

    \item \textbf{Flow-S}: An STFT-domain baseline following the STFT configuration and amplitude compression strategy in~\cite{pretrainSR}.

    \item \textbf{Flow-A}: An acoustic-only variant that takes noisy acoustic representations as input.

    \item \textbf{Flow-P}: A phonetic-only variant that directly models the distribution of phonetic representations, generating enhanced phonetic representations from noisy phonetic inputs.

    \item \textbf{Flow-M-P}: A Mel-domain variant conditioned on phonetic representations, designed to verify whether spectral-domain modeling with SSL conditioning underperforms entirely SSL-domain modeling, as adopted in PhASE-Flow.
\end{itemize}

All variants share the same architecture as PhASE-Flow, with input dimensions adjusted for each representation. Note that for the Mel, acoustic, and phonetic domains, the waveform is reconstructed using vocoders of identical architecture, trained independently for each domain and denoted as \textbf{Vocoder-M}, \textbf{Vocoder-A}, and \textbf{Vocoder-P}, respectively.

\subsection{Implementation Details}

For the encoder, we use the pre-trained WavLM-Large checkpoint. The vocoder adopts the Vocos backbone, comprising a linear projection into a 768-dimensional latent space, an attention module, and 12 ConvNeXt blocks~\cite{ConvNeXt} with an intermediate dimension of 2304. Final waveform reconstruction is performed via iSTFT (FFT size 1280, hop length 320). Following FlowSE, the DiT is configured with 22 layers, 16 attention heads, a 1024-dimensional hidden size, and a 2048-dimensional feed-forward network (FFN).

All models are trained on four NVIDIA RTX 4090 GPUs using the AdamW optimizer for 100k iterations with a total batch size of 128. The learning rate follows a linear warm-up to a peak of $5 \times 10^{-4}$ over the first 10\% of training steps, followed by cosine annealing to $1 \times 10^{-6}$. To ensure evaluation consistency, all FM-based models employ a 4-step ODE solver during inference.

\subsection{Experimental Results}

\subsubsection{Ablation Studies}

We first evaluate the analysis-synthesis performance of the three vocoders on the DNS 2020 clean speech set, with results shown in Table~\ref{tab:ablation_vocoder}. The vocoder based on acoustic representations (Vocoder-A) exhibits a clear advantage in UTMOS and SpkSim, which we attribute to the preservation of fine-grained acoustic details essential for high-fidelity reconstruction. In contrast, the vocoder based on phonetic representations (Vocoder-P) yields inferior performance in SpkSim, as it primarily captures high-level linguistic content at the expense of acoustic details and speaker identity.

We then compare the representation capabilities among the three vocoder-required baselines: Flow-M, Flow-A, and Flow-P. As shown in Table~\ref{tab:ablation_domain}, while Flow-P yields favorable non-intrusive metrics and dWER, its poor SpkSim hinders its practical application. Setting aside Flow-P, Flow-A consistently outperforms Flow-M on the no-reverb set. This supports our hypothesis that a disentangled, well-structured SSL manifold is more conducive to modeling clean speech than the Mel-spectrogram domain. Furthermore, vocoder-required baselines noticeably outperform the vocoder-free Flow-S in DNSMOS and UTMOS, suggesting that the pre-trained neural vocoder serves as a strong generative prior that explicitly enhances perceptual quality.

Regarding the phonetic conditioning effectiveness, PhASE-Flow yields consistent improvements across all metrics over Flow-A. In contrast, Flow-M-P reveals that phonetic guidance in the Mel domain only enhances SBS, LPS, and dWER, without yielding the broad gains observed in the SSL domain. These results substantiate that PhASE-Flow more effectively exploits the supplementary linguistic information.
By aligning semantics and acoustics within the SSL space, our framework mitigates representational mismatch, yielding a structured, coherent manifold for FM-based SE.

\subsubsection{Comparison on DNS 2020 no-reverb Test Set}

The left half of Table~\ref{tab:dns1} compares the performance of PhASE-Flow against the baselines on the no-reverb subset. As the sole discriminative baseline, TF-GridNet excels in preserving speaker identity and linguistic fidelity, achieving the highest SpkSim of 0.96 and a low dWER of 2.86\%. In contrast, generative approaches like StoRM, LLaSE-G1, and AnyEnhance exhibit noticeable linguistic degradation. Their dWER even exceeds that of the unprocessed noisy speech, indicating a high susceptibility to hallucinations.
Notably, TF-GridNet also achieves competitive non-intrusive scores, which we attribute to the high-SNR nature of the no-reverb set—a condition inherently favoring discriminative models.

Compared to all baselines, PhASE-Flow demonstrates a remarkably balanced and robust performance. For non-intrusive metrics, it yields DNSMOS scores comparable to LLaSE-G1 and AnyEnhance, while achieving a clear superiority in UTMOS (4.11). On representation-based metrics, PhASE-Flow achieves the highest SBS of 0.93 and LPS of 0.97. Its SpkSim is marginally lower than that of TF-GridNet, confirming its effectiveness in preserving speaker characteristics. Furthermore, achieving the lowest dWER of 2.79\% proves that our approach effectively mitigates the hallucination artifacts typically associated with generative models.

\subsubsection{Comparison on DNS 2020 with-reverb Test Set}

The right half of Table~\ref{tab:dns1} shows the results on the with-reverb subset. PhASE-Flow maintains robust performance, delivering the highest non-intrusive scores and consistently leading in LPS and SBS. We attribute the general degradation of SpkSim and dWER in generative models—often worse than noisy speech—to hallucinations exacerbated by complex reverberation. Nevertheless, PhASE-Flow significantly outperforms all other generative baselines with a SpkSim of 0.75 and a dWER of 13.19\%, trailing only TF-GridNet. These results confirm that carefully curated SSL features provide a promising representation manifold that preserves fine-grained acoustic details and speaker characteristics.

\section{Conclusion}

In this paper, we introduce PhASE-Flow, an FM-based SE framework that models speech distributions directly within the SSL domain. Our work demonstrates that this approach offers a robust and well-structured alternative to conventional spectral-domain methods. Experiments show that PhASE-Flow achieves superior perceptual quality and speaker similarity while effectively suppressing the hallucinations typical of generative models. Furthermore, the framework remains highly efficient, delivering competitive performance with only four sampling steps. Future work will extend PhASE-Flow to broader speech tasks and optimize it for real-time deployment.

\section{Acknowledgments}

This work was supported by the National Natural Science Foundation of China (Grant No. 12274221) and the Yangtze River Delta Science and Technology Innovation Community Joint Research Project (Grant No. 2024CSJGG1100).

\section{Generative AI Use Disclosure}

The authors confirm that no generative AI tools were used to create any original ideas, analyses, or substantial content in this manuscript. Generative AI was employed exclusively for minor language editing and polishing to enhance clarity and readability. The authors assume full responsibility and accountability for the integrity, accuracy, and content of the work presented herein.


\bibliographystyle{IEEEtran}
\bibliography{mybib}

\end{document}